\documentstyle[12pt]{article}
\begin{document}
\baselineskip 0.85cm
\begin{center}
\baselineskip 0.85cm
\vskip 0.4in
{\Large \bf Comment on ``Quasiparticle Decay Effects in the
Superconducting Density of States: Evidence for d-Wave Pairing
in the Cuprates''}
\vskip 2.0in
{\bf
Georgios VARELOGIANNIS}
\vskip 0.11in
{\em Dipartimento di Fisica\\
     Universit\`a degli studi di Roma "La Sapienza"\\
     Ple. A.Moro 2, 00185 Roma, Italy}
\vskip 0.7in
\end{center}
\newpage

In a recent letter [1], D.Coffey and L.Coffey analyzed some
characteristic anomalous structures in tunneling
and photoemission spectra of cuprates. They concluded that the
dip-like structure which appears above the gap
might be considered
as evidence for d-wave pairing.
In the analysis of [1], this structure is due
to ``deviations from weak coupling mean-field behavior
of the superconductivity in those materials''. While we perfectly
agree with this statement, we disagree with the
analysis and conclusions of [1] as well with the supposed origin
of the deviations from mean-field behavior.

When quasiparticle lifetime effects become important,
for the descritption of superconductivity it is necessary
to consider a retarded
strong coupling approach.
The necessity of a strong coupling approach is valid even if the
symmetry of the pairing is d-wave [2]. The errors in the analysis
and conclusions
of [1] are due to the use of a non-retarded framework, which
should exclude quantitative comparison with experiment.
In the case of d-wave pairing for example, the use of a retarded formalism
led to the reduction of the critical temperature by an order
of magnitude [2] compared to that obtained by a non-retarded approach.

The arguments advanced in [1] related with a systematic
localization of the dip at $3\Delta$ in SIS experiments are irrelevant
first because $\Delta$ is not known with sufficient precision
in high-$T_c$ materials and the imprecision on the energetic position
of the dip is even larger, secondly because a non-retarded approach
as that used in [1] cannot give relevant quantitative information and
thirdly because they disagree with the ARPES data where the dip
is also observed [3].

The analysis of Ref. [1] is in fact not only quantitatively
but also qualitatively \underline{contradictory} to the
ARPES data of the Stanford group [3] where an anisotropic
gap supporting a d-wave scenario is reported. In fact within the analysis
of Coffeys [1] the dip-like structure should appear in the direction
where the gap is absent in a d-wave scenario
while \underline{no dip structure} should appear
in the direction where the gap is present. Clearly within the analysis of
Ref. [1] the dip is related to the {\it absence} of the gap. But in the
photoemssion data of the Stanford group (and in all the other
photoemission data) exactly
the inverse happens. In the data of Ref. [3] {\it the dip structure
appears in the direction where the gap is present while no
dip appears in the direction where the gap is absent}.
Therefore when arguing in support of the d-wave ideas, it is necessary
to choose between Refs. [1] and [3] since they are contradictory.
If the analysis of [1] is valid, then ARPES data exclude
the d-wave hypothesis since no dip is seen in the direction
in which the gap is absent.

Recent extensions of calculations [4] of
the density of states of excitations within conventional s-wave
Eliashberg theory to
higher frequencies, reported an anomalous dip-like structure
at temperature independent frequencies followed by a
second peak or broad band [5]. Not only the form but also the
temperature dependence of these last structures [5] are very similar
to the experimental structures discussed in [1].
The exact frequency position (in $\Delta$ units) of the dip-like structure
is coupling strength dependent [5], and could appear
at $3\Delta$ in some SIS experiments.
Those structures are \underline{not} associated with the characteristic
frequencies of the boson mediators but are
due to the breakdown of Fermi liquid picture for the
virtually excited states occupied by the paired electrons
because of strong electron-phonon (or other boson) coupling [5].
Within this analysis one obtain a natural explanation to the
fact that the dip is more visible in ARPES experiments in the
directions in which the gap is larger [6]. In addition,
the recently reported asymetry in SIN tunnel data
where the dip is visible only at negative sample bias [7]
excludes the anlysis of Ref. [1] and can be naturally understood [6]
within the s-wave strong coupling analysis [5].

The whole anomalous behavior of the density of states
(finite density of states inside the gap, anomalous T dependence of the
gap, dip and second peak structures)
indicates that cuprates are at the beginning of a cross-over
from BCS superconductivity to Bose condensation [5].
This happens when the gap becomes comparable to the phonon energies
(and this is probably the case in cuprates and fullerides [8]),
and reflects the physical constraint that the distance the paired electron
covers during the absorbtion of the virtual phonon cannot be larger than
the superconducting coherence length.
The dip structure
should be a characteristic of all the materials that in the
analysis given in Ref. [9] of the Uemura plot, are close to the
cross-over regime (high-$T_c$ cuprates
and fullerides, heavy fermions etc.) [10].

Notice that
the observation, after the publication of Ref. [1],
of an analogous dip-like structure
in fullerides [9],
excludes the analysis of [1],
since for obvious reasons
nowhere in the litterature
a d-wave scenario has been proposed for fullerides.
%n addition, although the traduction of the binding energy spectra
%rovided by photoemission in terms of $\Delta$ units is not obvious,
%o our analysis, the dip in the data of Ref. [11] are in good agreement
%ith results of photoemission in Cuprates and the analysis of Ref. [5].
The eventual observation of the dip structure in heavy fermion compounds
will further confirm the analysis of Ref. [5].

In conclusion it is obvious that the dip-like structure
\underline{cannot} be
considered as evidence for d-wave pairing.

\newpage
\centerline{\bf REFERENCES}
\begin{enumerate}

\item D. Coffey and L. Coffey, Phys. Rev. Lett {\bf 70},
1529 (1993)

\item P. Monthoux and D. Pines, Phys. Rev. B {\bf 47},
6069 (1993)

\item D.S. Dessau et al., Phys. Rev. Lett. {\bf 66}, 2160 (1991)

\item P.B. Allen and D. Rainer, Nature {\bf 349}, 396 (1991)

\item G. Varelogiannis, Phys. Rev. B {\bf 51}, 1381 (1995)

\item G. Varelogiannis et al., Preprint cond-mat/9507052

\item Ch. Renner and \O. Fischer, Phys. Rev. B {\bf 51}, 9208 (1995)

\item G. Varelogiannis, Phys. Rev. B {\bf 50}, 15974 (1994)

\item F. Pistolesi and G.C.Srtinati, Phys. Rev. B {\bf 49}, 6356 (1994)

\item G. Varelogiannis and L.Pietronero, Preprint

\item M. Knupfer et al., Phys. Rev. B {\bf 47}, 13944 (1993)

\end{enumerate}

\end{document}